\documentclass[prl,preprint,superscriptaddress,floatfix]{revtex4}

\usepackage{graphicx}
\usepackage{dcolumn}

\begin{document}

%\addtolength{\topmargin}{2cm}

\title{Tackling the Fermionic Sign Problem in the Auxiliary-Field Monte Carlo Method}

\author{G.~Stoitcheva}
\author{ W. E.~Ormand}
\affiliation{Lawrence Livermore National Laboratory, P.O. Box 808, L-414, Livermore, CA 94551, USA} 
\author{D. Neuhauser}
\affiliation{Department of Chemistry and Biochemistry, University of California, Los Angeles, CA 90024, USA} 
\author{D. J. Dean}
\affiliation{Physics Division, Oak Ridge National Laboratory, Oak Ridge, TN 37831, USA} 
\begin{abstract}
We explore a novel and straightforward solution to the sign problem that has plagued the Auxiliary-field Monte Carlo (AFMC) method applied to many-body systems for more than a decade. 
We present a solution to the sign problem that has plagued the Auxiliary-field Monte Carlo (AFMC) method for more than a decade and report a breakthrough where excellent agreement between AFMC and exact CI calculations for fully realistic nuclear applications is achieved. This result offers the capability, unmatched by other methods, to achieve exact solutions for large-scale quantum many-body systems.
\end{abstract}
\maketitle

\narrowtext

The quantum many-body problem is the foundation of much of modern physics and chemistry, and one of the great challenges in theoretical physics is to develop a fully microscopic solution that includes the full range of quantum correlations.
% that includes the full range of quantum correlations. 
Traditionally, configuration-interaction (CI) methods, such as the nuclear shell model, were a method of choice. Valence particles occupy a set of single-particle orbitals and influence each other through an effective interaction. In this valence space is a set of $N_D$ basis states, $\phi_i$, used to construct the eigenstates of the Hamiltonian $\hat{H}$.
% i.e., $\Psi_\mu=\sum_i \alpha_{\mu i}\phi_i$. 
CI reduces to a matrix-diagonalization problem by finding the eigenvalues of the matrix $H_{ij}=\langle\phi_i | \hat{H} | \phi_j\rangle$. While powerful, CI is a brute-force method facing substantial computational challenges. Diagonalization methods for sparse matrices typically scale as $N_D^{1.25}$ and current limits are $N_D\sim10^{9-10}$. But, the number of basis states increases dramatically with particle number. For two-component systems, such as nuclei, the number of basis states increases as 
\begin{equation}
N_D \approx \left( \begin{array}{c} N_s^p \\ N_v^p \end{array} \right) \left( \begin{array}{c} N_s^n \\ N_v^n \end{array}\right) ,
\label{number-states}
\end{equation}
where $N_s^{p(n)}$ is the number of proton (neutron) single-particle states in the configuration space, and $N_v^{p(n)}$ is the number of proton (neutron) valence particles. Consequently, for mid-mass nuclei, where $N_s^{p(n)} \sim 40$ and $N_v^{p(n)}\sim20$ would be typical, the matrix dimension would be of the order $10^{20}$, which to solve would require a computer $10^{12}$ times more powerful than any available today.

Monte Carlo methods offer an attractive alternative to CI as their computational effort scales more gently with particle number. Indeed, Monte Carlo methods have been applied to a wide 
variety of many-fermion problems in physics and chemistry; with applications in condensed matter, nuclear structure, and lattice quantum chromodynamics (see Ref.~\cite{QMC_methods}). Unfortunately, Monte Carlo methods applied to fermionic systems generally suffer from the well-known sign problem (where the sampling weight function is not positive definite), which substantially limits their efficacy. 
Here, we will address the sign problem with the Auxiliary-Field Monte Carlo (AFMC) method~\cite{AFMC_0} based on the Hubbard-Stratonovich (HS) transformation~\cite{HS}. AFMC has had wide applications; including the Hubbard model~\cite{AFMC_Hubbard} and nuclei~\cite{AFMC_1, AFMC_2}. The full power of the AFMC method has not been realized because of the sign-problem, and past nuclear physics studies have been limited to using semi-realistic interactions with good sign or an extrapolation method based on breaking up the Hamiltonian into ``good-sign'' and ``bad-sign'' parts~\cite{YA}. Here, we report a breakthrough solution to the sign problem based on shifting the contour as initially proposed in Ref.~\cite{Rom}. We show that the shifted-contour method is practical and extraordinarily effective in mitigating the sign problem for fully realistic Hamiltonians in nuclear systems. These results offer a clear pathway to perform large-scale many-body calculations that include the full range of quantum correlations.

The AFMC method for generic rotationally Hamiltonians is outlined in detail in Ref.~\cite{AFMC_2}, and here we present the central features germane to our solution to the sign problem. AFMC is based on the imaginary-time evolution operator $e^{-\beta\hat H}$ to either filter from an arbitrary trial wave function, $\phi_0$, the ground-state (GS) value for the operator $\hat{\Omega}$ via
\begin{equation}
\langle \hat{\Omega}\rangle_{GS} = \lim_{\beta \rightarrow \infty} \frac{
\langle\phi_0 | e^{-\beta\hat{H}/2}\hat{\Omega}e^{-\beta\hat{H}/2} | \phi_0 \rangle}
{\langle\phi_0 | e^{-\beta\hat{H}} | \phi_0 \rangle}\;,
\label{zero_temp}
\end{equation}
or to compute the thermal expectation value $\langle \hat{\Omega}\rangle_\beta$ 
\begin{equation}
\langle \hat{\Omega}\rangle_\beta =\mathrm{Tr}_{(Z,N)}\left[e^{-\beta \hat{H}}\hat{%
\Omega}\right] / \mathrm{Tr}_{(Z,N)}\left[e^{-\beta \hat{H}}\right]\;,  \label{thermal}
\end{equation}
where $\mathrm{Tr}_{(Z,N)}$ denotes the $Z$-proton and $N$-neutron projected trace.
Eqs.~(\ref{zero_temp}) and (\ref{thermal}) are distinct and complementary approaches. Along with GS properties, the thermal formalism permits us to calculate structure information, such as electro-weak transition strengths, at finite temperature and is the optimal procedure for computing the density of states~\cite{Levit}; a critical ingredient needed to describe astrophysical nucleosynthesis processes~\cite{nucleosynthesis}.
%, such as the s-, r-, and rp-processes~\cite{nucleosynthesis}. 
%Without a fully microscopic theory, the density of states is typically described with the backshifted
%Fermi-gas model of Gilbert and Cameron~\cite{GilbertCameron}, whose empirical parameters are one
%of the largest sources of uncertainty in determining the reaction rates required for large-scale
%nucleosynthesis networks~\cite{nucleosynthesis,ErrorRates}.  
On the other hand, the ``zero-temperature'' formalism, Eq.~(\ref{zero_temp}), is an efficient way to compute GS observables. 

Since any 
two-body Hamiltonian may be written in quadratic form as
\begin{equation}
\hat{H}=\sum_{\alpha }\varepsilon_{\alpha}\hat{\Theta}_{\alpha}
+\frac{1}{2}\sum_{\alpha}V_{\alpha }\hat{\Theta}^2_{\alpha}\;, 
\label{Ham}
\end{equation}
where here we choose $\hat{\Theta}_{\alpha }$ to be a generalized one-body density operator, $V _{\alpha}$ the strength of the
two-body interaction, and ${\varepsilon }_{\alpha}$
the single-particle energies, we simplify $\exp(-\beta\hat H)$ making use of 
the (HS) transformation~\cite{HS}
\begin{equation}
\mathop e\nolimits^{\frac{1}{2}\Lambda\hat{\Theta} ^{2}} = \sqrt{\frac{
{\left|\Lambda \right| }}{{2\pi }}}\int {d\sigma \mathop e\nolimits^{-\frac{1}{2}%
\left| \Lambda\right|\sigma ^{2}+s\sigma \Lambda\hat{\Theta} }},
\label{HS_trans}
\end{equation}
where $s=\pm 1$ if 
$\Lambda \geq 0$ or $\pm i$ if $\Lambda < 0$ and $\sigma$ is the associated auxiliary field. 
Setting $\Lambda=-\beta V_\alpha$, we have
%the exponential of the full Hamiltonian may then be written as multi-dimensional integral
\begin{equation}
\mathop e\nolimits^{-\beta \hat{H}}=
\int \mathcal{D}[\vec \sigma]  G(\vec\sigma)
\mathop e\nolimits^{-\beta \hat{h}(\vec \sigma)},
\label{eq_22}
\end{equation}
where $G(\vec\sigma)=\exp(-\frac{1}{2}\beta\sum_\alpha \left| V_{\alpha}\right| \sigma_{\alpha} ^{2})$ is the Gaussian factor, the volume element is 
$\mathcal{D}[\vec \sigma]=\prod\nolimits_{\alpha }{d\sigma _{\alpha}}\sqrt{ 
\beta \mid V_\alpha\mid/2\pi}$, and  
$\hat{h}(\vec\sigma)=\sum_\alpha{(\varepsilon_{\alpha} }+s_{\alpha}
V_{\alpha} \sigma_{\alpha} )\hat{\Theta}_{\alpha}$. 
Since in general the operators $\hat\Theta_\alpha$ do not commute,  we split $e^{-\beta \hat H}$ into 
$N_{t}$ time-slices, i.e.,
$e^{-\beta \hat{H}}=[ e^{-\Delta \beta \hat{H}}] ^{N_{t}}$ (all calculations presented here are with $\Delta\beta=1/32$~MeV$^{-1}$) and apply the HS transformation at each time slice. Eqs.~(\ref{zero_temp})-(\ref{thermal}) can then be written as
\begin{equation}
\langle {\hat{\Omega}}\rangle =\frac{{\int {\mathcal{D}[\vec \sigma ]W(\vec\sigma )
\langle \hat\Omega \rangle _{\vec\sigma }}}}
{{{\int {D[\vec\sigma]W(\vec\sigma )}}}}.
\label{observe_integral}
\end{equation}
Defining the one-body imaginary-time propagator as
\begin{equation}
U_\sigma=e^{-\Delta\beta h(\vec\sigma(N_t))}\cdots 
e^{-\Delta\beta h(\vec\sigma(1))}
\end{equation}
we express the weight function $W(\vec\sigma)$ and $\langle \hat \Omega \rangle _{\vec\sigma }$ as
\begin{equation}
W(\vec\sigma)=G(\vec\sigma)\mathrm{Tr}_{(Z,N)}\left[U_\sigma\right],~~
\langle \hat\Omega \rangle _{\vec\sigma }=\frac{ \mathrm{Tr}_{(Z,N)}\left[\hat \Omega U_\sigma\right]}
{\mathrm{Tr}_{(Z,N)}\left[U_\sigma\right]}.
\label{W_obs}
\end{equation}
%we express the weight function $W(\vec\sigma)$ as
%\begin{equation}
%W(\vec\sigma)=G(\vec\sigma)\mathrm{Tr}_A\left[U_\sigma\right],
%\label{weight}
%\end{equation}
%and $\langle \hat \Omega \rangle _{\vec\sigma }$ as
%\begin{equation}
%\langle \hat\Omega \rangle _{\vec\sigma }=\frac{ \mathrm{Tr}_A\left[\hat \Omega U_\sigma\right]}
%{\mathrm{Tr}_A\left[U_\sigma\right]}.
%\label{observe}
%\end{equation}

The auxiliary fields $\sigma_\alpha$ are not just parameters introduced for numerical convenience, but have physical significance. Their presence in $\hat h(\vec\sigma)$ essentially constructs a constrained mean field. Indeed,  the maximum of the weight function, $W(\vec\sigma)$, corresponds to the Hartree mean-field solution~\cite{AFMC_2}, satisfying the self-consistent condition
\begin{equation}
\sigma^{MF}_\alpha=-s_\alpha \mathrm{sgn}(V_\alpha)\langle \hat\Theta_\alpha\rangle_{\vec\sigma^{MF}}.
\label{mean_field}
\end{equation}

The principal advantage of the AFMC is that overall the computational effort scales much more gently with particle number. For example, the number of proton (neutron) auxiliary fields is at most $(N^{p(n)}_s)^2\times N_t$. Thus, while for the case where $N^p_s=N^n_s=40$ and $N_v^p=N_v^n=20$ conventional CI methods are confronted with matrices  with dimension $\sim 10^{20}$, the number of AFMC fields with $N_t=100$ is 320,000. 

Given the large number of auxiliary fields, Eq.~(\ref{observe_integral}) is evaluated using Monte Carlo methods. Thus,
%we evaluate Eq.~(\ref{observe_integral}) using Monte Carlo methods. Thus, 
\begin{equation}
\langle \hat \Omega \rangle_{MC} = 
\frac{1}{N} \sum_i \langle \hat \Omega \rangle _{\vec\sigma_i },
\label{Monte-Carlo}
\end{equation}
where $N$ is the number of samples (typically 4000), and $\vec\sigma_i$ is distributed according to $W(\vec\sigma)$. The uncertainty in the integral is then governed by the variance. Central to the Monte Carlo evaluation is that $W(\vec\sigma)$ be positive definite. For rotationally invariant applications, the general conditions for which 
$W(\vec\sigma)\ge 0$ was examined in Ref.~\cite{AFMC_2}, and was found to be true only for a small class of semi-realistic interactions, such as pairing-plus-quadrupole, for even-particle systems.  Without a positive-definite weight function, we can try to proceed by sampling with $\left| W(\vec\sigma)\right|$. Eq.~(\ref{Monte-Carlo}) is modified by the presence of the ``sign'', $\Phi(\vec\sigma_i)=W(\vec\sigma_i)/\left| W(\vec\sigma_i)\right|$, multiplying $\langle \hat \Omega \rangle _{\vec\sigma_i }$, and is normalized by the average sign $\langle\Phi\rangle$.
%$\langle\Phi\rangle=\frac{1}{N}\sum_i \Phi(\vec\sigma_i)$. 
Fig. 1 demonstrates how AFMC fails for general Hamiltonians. The figure shows the thermal energy as a function of $\beta$ for $^{28}$Mg within the $sd$-shell using the realistic Hamiltonian of Wildenthal~\cite{Wildenthal}.
The solid line shows the exact CI result, where all 28,503 eigenvalues were obtained with OXBASH~\cite{OXBASH}. The circles show the AFMC calculation while sampling $\left|W(\vec\sigma)\right|$ with the Metropolis algorithm~\cite{Metropolis}. In general, sampling with $\left| W(\vec\sigma)\right|$ breaks down for $\beta \ge 0.4$ MeV$^{-1}$.
%%%%%%%%%%%%%%%%%%%%%%%%%%%%%
% FIGURE 1
%%%%%%%%%%%%%%%%%%%%%%%%%%%%%%%%%
\begin{figure}[htb]
\centering
 \includegraphics[width=4.0in]{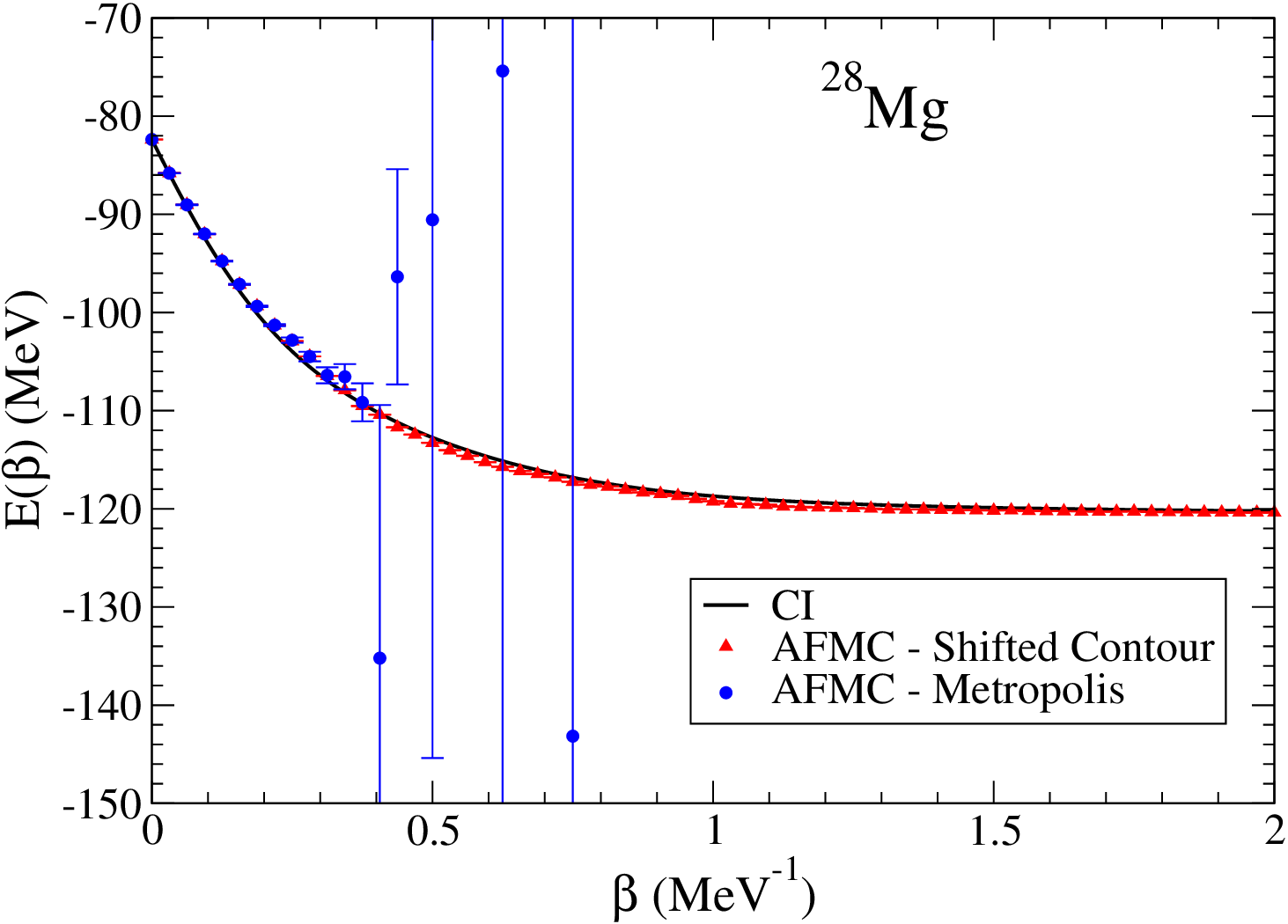}
% \includegraphics[width=4.0in]{Ormand_fig_1.jpg}
% \includegraphics[width=2.9in]{Ormand_fig_1.jpg}
%\vspace{-0.15in}
%\hspace{-1.4in}
\caption{Thermal energy for the nucleus $^{28}$Mg computed with the $sd$-shell Hamiltonian of Wildenthal. The solid line shows the exact CI result obtained from all 28,503 shell-model eigenvalues. The (blue) circles show the AFMC result using Metropolis sampling on $\left| W(\vec\sigma)\right|$. The (red) triangles show the results obtained using the shifted-contour method. }
\label{shell-m}
\end{figure}

To address the sign problem, we rewrite the two-body Hamiltonian as
\begin{eqnarray}
\sum_\alpha V_\alpha\hat\Theta^2_\alpha =
\sum_{\alpha}V_{\alpha }({\hat\Theta}_\alpha-\bar\sigma_\alpha)^2
+V_{\alpha }(2\hat{\Theta}_{\alpha}\bar\sigma_\alpha-
\bar{\sigma}_\alpha^2), \label{eq_3}
\end{eqnarray}
 and apply the HS transformation to the quadratic $({\hat\Theta}-\bar\sigma)^2$ terms, giving for $e^{-\Delta\beta \hat H}$
\begin{equation}
\int \mathcal{D}[\vec\sigma] e^{-\frac{1}{2}\Delta\beta\sum_\alpha |V_\alpha |\sigma^2_\alpha 
-V_\alpha (2s_\alpha\sigma_\alpha\bar\sigma_\alpha+\bar\sigma^2_\alpha)} e^{-\Delta\beta \hat h(\vec\sigma)},
\label{shift}
\end{equation}
where now $\hat{h}(\vec\sigma)=\sum_\alpha{[\varepsilon_{\alpha} }+
V_{\alpha} (s_{\alpha}\sigma_{\alpha} +\bar\sigma_\alpha)]\hat{\Theta}_{\alpha}$.
With the shift $\bar\sigma_\alpha$, the maximum of the weight function is now
\begin{equation}
\sigma_\alpha=-s_\alpha \mathrm{sgn}(V_\alpha)(
\langle \hat\Theta_\alpha\rangle_{\vec\sigma}-\bar\sigma_\alpha).
\label{maximum}
\end{equation}
Thus, if we choose $\bar\sigma_\alpha=\sigma^{MF}_\alpha$, the maximum of the weight function occurs at $\sigma_\alpha=0$.  The presence of $\bar\sigma_\alpha$ in the exponential factors in Eq.~(\ref{shift}) is important. 
For $V_\alpha < 0$, $W(\vec\sigma)$ is shifted to the origin. While for $V_\alpha > 0$, the overall maximum is shifted into the complex plane with the maximum along the real axis at $\sigma_\alpha=0$. Further, a static phase is introduced that suppresses the bad sign as we sample the along the real axis. 
%%%%%%%%%%%%%%%%%%%%%%%%%%%%%%%%
% FIGURE 2
%%%%%%%%%%%%%%%%%%%%%%%%%%%%%%%%%
\begin{figure}[htb]
\centering
\includegraphics[width=3.5in]{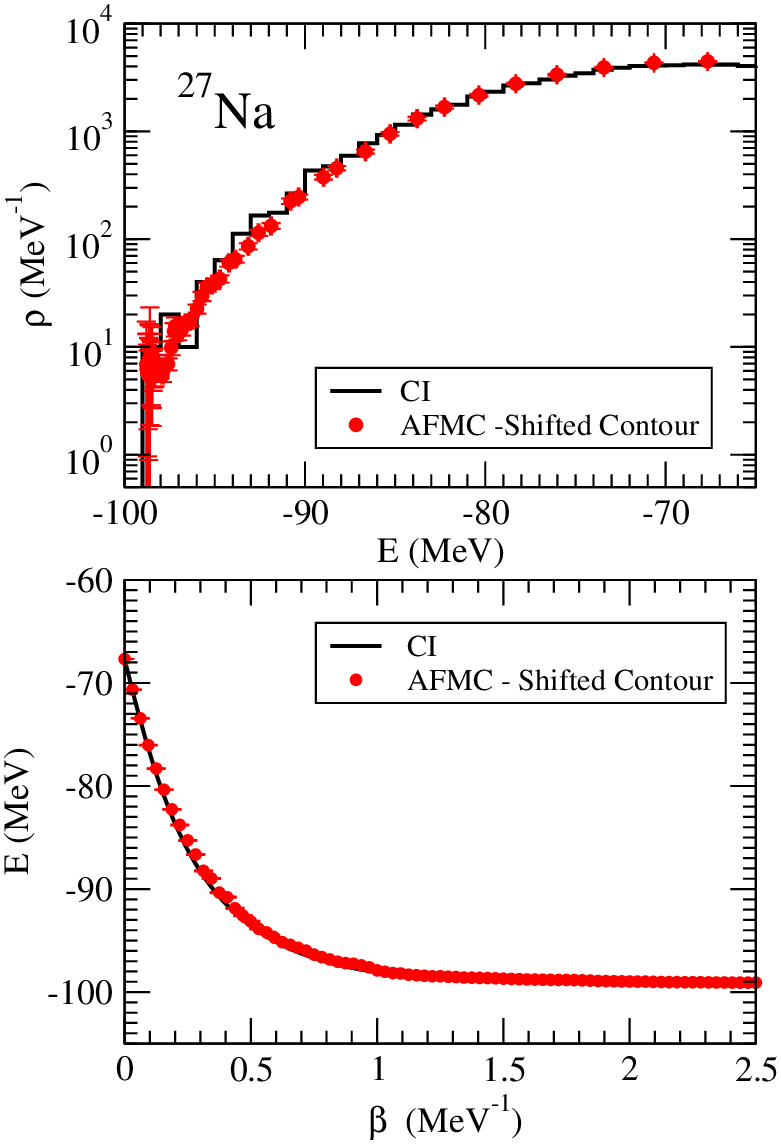}
\vspace{-0.10in}
%\hspace{2.4in}
\caption{Thermal energy and the state density $\rho(E)$, for the nucleus $^{27}$Na computed with the $sd$-shell Hamiltonian of Wildenthal. The solid line shows the exact CI result obtained from the eigenvalues. The circles show the AFMC result using the shifted-contour method. }
\label{na27}
\end{figure}

We note that since the maximum of $W(\vec\sigma)$ is centered about $\sigma_\alpha=0$ we can Monte Carlo sample the $\sigma$-fields with the overall Gaussian factor 
$G(\vec\sigma)$. The advantage of sampling with the Gaussian factor is that it offers an efficient method to sample uncorrelated values $\vec\sigma_i$.
In Fig.~\ref{shell-m}, the results of AFMC calculation of the thermal energy for $^{28}$Mg using the shifted-contour method with Gaussian sampling is shown (triangles) and compared to the exact CI result as well as with Metropolis sampling on $|W(\vec\sigma)|$. Shifting the contour yields agreement with the exact thermal calculation, which clearly represents a significant improvement over previous capability. With the zero-temperature formalism, at $\beta=3.0$ MeV$^{-1}$ we compute a GS energy of -120.370(25) MeV, which is in good agreement with the CI result of -120.532 MeV. 

In Fig.~\ref{na27}, we compare the CI and AFMC results for the thermal energy and the 
state density, $\rho(E)$, for  
$^{27}$Na in the $sd$-shell using the Wildenthal interaction.  With the zero-temperature formalism,  we obtain the GS energy of -99.106(55) MeV, which is also in good agreement with the exact result of -99.230 MeV.
The state density (the total density of states including the $(2J+1)$ degeneracy for each state of angular momentum $J$) can be computed with the saddle-point approximation for the inverse Laplace transform of the partition function, i.e.,
\begin{equation}
\rho(E)=e^{\ln Z(\beta)+\beta E(\beta)}/\sqrt{-2\pi\partial E(\beta)/\partial\beta},
\label{density}
\end{equation}
where $\ln Z(\beta)=-\int_0^\beta d\beta^\prime E(\beta^\prime)+\ln Z(0)$, and $Z(0)$ is the total number of states given by Eq.~(\ref{number-states}). These $^{27}$Na results are significant because not only is excellent agreement with the CI calculation achieved, but beforehand odd-systems suffered from the sign problem even for the semi-realistic, good-sign interactions identified in Ref.~\cite{AFMC_2}. 
%%%%%%%%%%%%%%%%%%%%%%%%%%%%%%%%
% FIGURE 3
%%%%%%%%%%%%%%%%%%%%%%%%%%%%%%%%%
\begin{figure}[htb]
\centering
\includegraphics[width=3.5in]{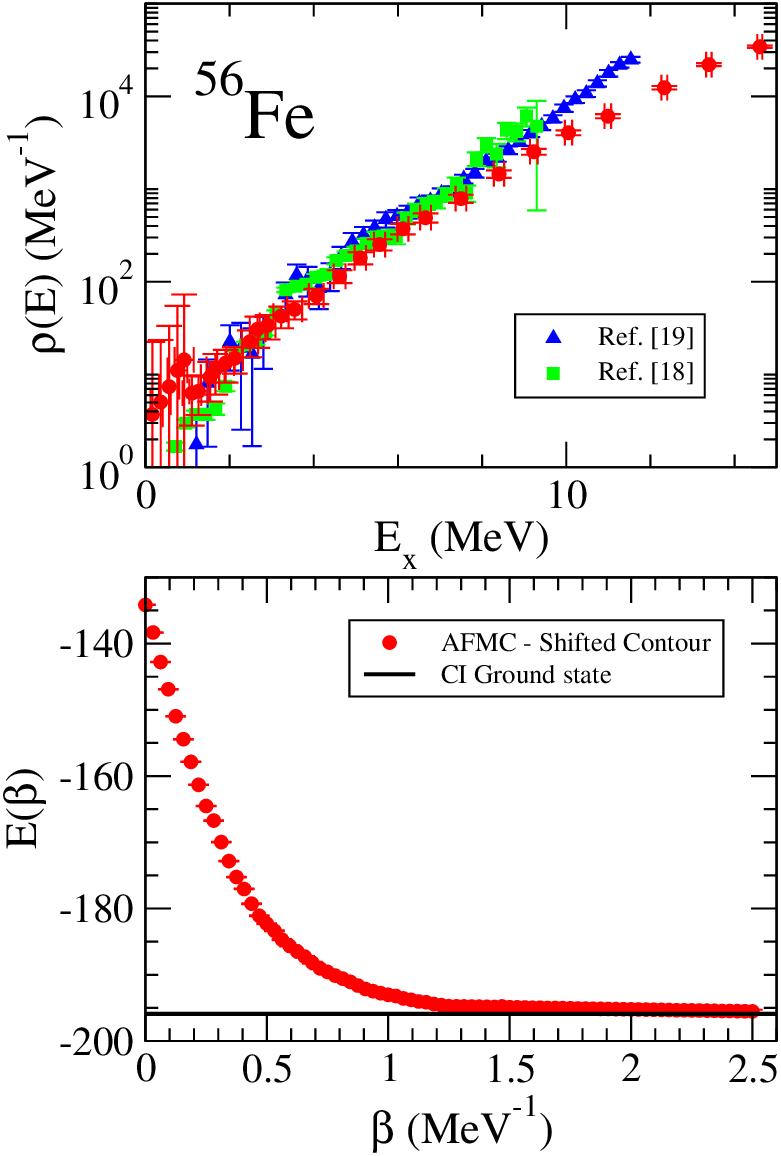}
\vspace{-0.10in}
%\hspace{2.4in}
\caption{Thermal energy and the state density $\rho(E)$ \ for the nucleus $^{56}$Fe computed with the GXPF1A $fp$-shell Hamiltonian. The solid line in the bottom panel shows the exact CI result for GS energy. The circles show the AFMC result using the shifted-contour method. In the upper panel, the calculated state is compared with values inferred from recent experiments [(squares)~\cite{LD1}, (triangles)~\cite{LD2}].}
\label{fe56}
\end{figure}

In Fig.~\ref{fe56}, we show results for the more challenging case of $^{56}$Fe, where the  GXPF1A interaction~\cite{GXFP1}  was used in an active model space comprised of the $0f-1p$ orbits. Here, $N_s^{p(n)}=20$, $N_v^p=6$, and $N_v^n=14$, and the number of CI basis states with $J_z=0$ is  $\approx$~501M. In this case, we obtained the GS energy with the shell-model code REDSTICK~\cite{REDSTICK}, which is represented by the solid line in the figure. The shifted-contour AFMC calculation is clearly converging to the full-space CI result.  With the zero-temperature formalism we calculate a GS energy of -195.687(107) MeV, which is in good agreement with the CI result of -195.901 MeV. The computational advantage of AFMC for large model spaces is evident as the zero-temperature calculation for $^{56}$Fe took 12 CPU hours~\cite{uP}, as opposed to 1000 CPU hours for CI~\cite{atlas}. In the upper panel, we compare the calculated state density 
%as a function of excitation energy 
with values inferred in recent experiments~\cite{LD1,LD2}. Overall agreement with the inferred experimental quantities is achieved. Here, our intent is to demonstrate a new capability, thus our AFMC calculation consists of just one major shell. Consequently, negative-parity and higher-lying states are outside this model space, and the calculated state density will under predict the observed state density at higher excitation energies. In principal, there are no underlying computational difficulties in extending our calculations to include more major shells; only the question of the appropriate effective interaction. 
%These results demonstrate a new capability where a fully microscopic description of the density of 
%states is now viable. 

We present a solution to the sign problem for the AFMC method applied to many-body systems based on shifting the quadratic part of the two-body Hamiltonian. The optimal choice for the shift is the fields associated with the Hartree mean-field solution for each specific value of $\beta$. This choice shifts the maximum of the integrand to the origin; permitting efficient sampling using the Gaussian factor. For bad sign components of the Hamiltonian, the shift introduces phases that mitigate the presence of negative signs in the weight function as the fields are sampled along the real axis. 
%By shifting the contour, the full potential of the AFMC method may now be realized; giving a capability 
%unmatched by other methods to perform large-scale many-body calculations.
% that include the full range of quantum correlations. 
With $\Delta\beta = 1/32$ MeV$^{-1}$,  the thermal energy is typically reproduced at the level of 300 keV or better, while the GS energies are reproduced to within 150-200 keV. This is a substantial improvement over previous attempts~\cite{YA}, where deviations of the order 1 MeV from CI results 
were common~\cite{caurier}. We note that this is generally the level of accuracy achieved by the effective interactions themselves~\cite{Wildenthal,GXFP1}. An exciting possibility for the future is to combine the AFMC with traditional mean-field approaches based on Skyrme-like interactions~\cite{skyrme} to develop a universal picture for nuclei.  In addition, given that the mean field is ubiquitous for many-body systems, it is likely that the shifted-contour method will have wide ranging applications to the quantum many-body problem across many subfields of theoretical physics and chemistry. 

{\bf Acknowledgments} 
G.S. acknowledges useful discussions with M. Stoitsov. This work was performed under the auspices of the U.S. Department of Energy by the University of California, Lawrence Livermore National Laboratory under Contract W-7405-Eng-48. Support from LDRD contract 06-LW-13 is acknowledged. Oak Ridge National Laboratory is managed by UT-Battelle, LLC under contract No. DE-AC05-00OR22725.


\begin{thebibliography}{99}
%
\bibitem{QMC_methods} {\sl Quantum Monte Carlo Methods in Physics and Chemistry}, Nato Science Series C, vol. 525, ed. M. P. Nightingale and C. J., Umrigar, (Springer, Berlin, 1999).
%
\bibitem{AFMC_0}
G. Sugiyama and S. E. Koonin, Phys. Ann. Phys. {\bf 168},1 (1986)
%
\bibitem{HS}
J. Hubbard, Phys. Rev. Lett. {\bf 3} (1959), 77; R. L. Stratonovich, Dokl. Akad. Nauk. S.S.S.R. {\bf 115}, 1097 (1957).
%
\bibitem{AFMC_Hubbard}
S. R. White, {\it et al.}, Phys. Rev. B {\bf 40}, 506 (1989).
%
\bibitem{AFMC_1}
C. W. Johnson, S. E. Koonin, G. H. Lang, and W. E. Ormand, Phys.
Rev. Lett. {\bf 69}, 3157 (1992).
%
\bibitem{AFMC_2}
G. H. Lang, C. W. Johnson, S. E. Koonin, and W. E. Ormand, Phys. Rev. C {\bf 48}, 1518 (1993); W. E. Ormand, Prog. Theo. Phys. Supp. No. {\bf 124}, 37 (1996); S. E. Koonin, D. J. Dean, and K. Langanke, Phys. Rep. {\bf 278}, 1 (1997).
%
\bibitem{YA}  
Y. Alhassid, D. J. Dean, S. E. Koonin, G. Lang, and W. E.
Ormand, Phys. Rev. Lett. {\bf 72}, 613 (1994).
%
\bibitem{Rom}  
N. Rom, D. M. Charutz, and D. Neuhauser, Chem. Phys. Lett. {\bf 270}, 382 (1997);
R. Baer, M. Head-Gordon, and D. Neuhauser, J. Chem. Phys. {\bf 109}, 6219 (1998); N. Rom, E. Fattal, A. K. Gupta, E. A. Carter, and D. Neuhauser, J. Chem. Phys. {\bf 109}, 8241 (1998).
%
\bibitem{Levit}
A. K. Kerman and S. Levit, Phys. Rev. C {\bf 24}, 1029 (1981).
%
\bibitem{nucleosynthesis}
J. J. Cowan, F.-K. Thielemann, and J. W. Truran, Phys. Rep. {\bf 208}, 267 (1991).
%
%\bibitem{GilbertCameron}
%A. Gilbert and A. G. W. Cameron, Can. J. Phys. {\bf 43}, 1446 (1965).
%
%\bibitem{ErrorRates}
%K. Langanke, F.-K. Thielemann, and M. Wiescher, Lecture Notes in Physics, 383 (2004); P. 
%Descouemont and T. Rauscher, Nucl. Phys. {\bf A777}, 137 (2006); T. Rauscher, F.-K. Thielemann,
% and K.-L. Kratz, Phys. Rev. C {\bf56}, 1613 (1997).
%
\bibitem{Wildenthal}
B. H. Wildenthal, in {\sl Progress in Particle and Nuclear Physics}, ed. D. H. Wilkinson, (Pergamon, Oxford, 1984), vol. 11, p. 5.
%
\bibitem{OXBASH}
W. D. M. Rae, A. Etchegoyen, and B. A. Brown, {\sl OXBASH, The Oxford-Buenos Aires-MSU shell-model code}, Michigan State University Cyclotron Laboratory report No. 524 (1988).
%
\bibitem{Metropolis}
N. Metropolis, A. Rosenbluth, M. Rosenbluth, A. Teller, and E. Teller, J. Chem. Phys. {\bf 21}, (1087) 1953.
%
\bibitem{GXFP1}
M. Honma, T. Otsuka, B. A. Brown, and T. Mizusaki, Phys. Rev. C {\bf 65}, 061301(R) (2002)
%
\bibitem{REDSTICK}
W. E. Ormand, C. W. Johnson, REDSTICK, version 3.5, UCRL-CODE-230640.
%
\bibitem{uP}
With a single 1.9 GHz Power5 CPU on uP at LLNL.
%
\bibitem{atlas}
With 100 2.4 GHz Opteron CPUs on Atlas at LLNL.
%
\bibitem{LD1}
A. Schiller {\it et al.}, Phys. Rev. C {\bf 68}, 054326 (2003).
%
\bibitem{LD2}
A. V. Voinov {\it et al.}, Phys. Rev. C {\bf 74}, 014314 (2006).
%
\bibitem{caurier}
E. Caurier, {\it et al.}, Phys. Rev. C {\bf 59}, 2033 (1999)
%
\bibitem{skyrme}
D. Vautherin and D. M. Brink, Phys. Rev. C {\bf 5}, 626 (1972).

\end{thebibliography}
\end{document}